\documentclass[12pt]{article}

\def\be{\begin{equation}}
\def\ee{\end{equation}}

\begin{document}
\titlepage

\vspace*{0.5cm}
\begin{center}
{\Large \bf Bose-Einstein correlation to measure the size of event of different types  }\\


\vspace*{1cm}

V.A. Schegelsky,
and M.G. Ryskin

\vspace*{0.5cm}
 Petersburg Nuclear Physics Institute, NRC Kurchatov Institute, 
Gatchina, St.~Petersburg, 188300, Russia \\
\end{center}

\begin{abstract}
Bose-Einstein correlations of identical hadrons produced in 
high-energy $pp$ collisions at the LHC is a good instrument to probe the 
size of the domain which emits these hadrons in different classes of 
events.  This provides an additional information on the dynamics of 
multiparticle production. In particular this way we may  measure the 
radius of the colour tube/string which create the secondary pions.
\end{abstract}
\vspace*{0.5cm}

\section{Introduction}
Identical particles correlation (BEC) may be used as a good instrument to measure the size of the region from which the secondaries radiated in one or another class of events~\cite{hbt,gfg,kop,bec}. Indeed, according to Bose-Einstein statistics we have to sum the amplitude, $M_a$, where the particle with momentum $p_1$ is emitted at the point $r_1$ and the particle with momentum $p_2$ is emitted at the point $r_2$ with that, $|M_b|=|M_a|$, corresponding to the permutation: $p_1$ is emitted at $r_2$ and $p_2$ - at $r_1$. The phase difference between this two amplitudes is $\exp{i\vec r\vec Q}$ with $\vec r=\vec r_1-\vec r_2$ and $\vec Q=\vec p_1-\vec p_2$. For a large $Q>>1/<r>$ the interference contribution is practically vanishes due to a strong oscillations during the integration over the $r_1$ and $r_2$ positions. However at a very small
$Q<<1/<r>$ the factor $\exp{i\vec r\vec Q}\simeq 1$ is close to unity and this fact doubles the cross section. By this reason we expect the peak at small $Q$ in the two particle inclusive cross section 
\be
\frac{d\sigma}{d^3p_1d^3p_2}\propto 2|M_a|^2(1+<e^{i\vec r\vec Q}>)\ .
\ee
The width of this peak in $Q$ is the measure of the mean size of the radiation region.\\
\section{Radius of source depends on \\ the detail structure of event}
At the beginning the size of high energy interaction was studied as the function of energy using the whole ensemble of secondaries without any selection of events. It was shown that the radius increases with energy. However studying the BEC in the interval $\sqrt s=$ 0.2 to 0.9 TeV the UA1 group had shown that for a fixed value of charged particle density the size of the emitting region is independent on
   $\sqrt{s}$ but increases with the  particle densities~\cite{ua1bec}.\\
In ~\cite{smrk} this was interpreted as the fact that actually the size of domain which emits the pions in the central rapidity region depends not on the energy but on the number of Pomerons exchanged between the colliding protons.
Strictly speaking we do not know from this experiment whether this were the Pomerons or some other objects of a relatively small transverse size (some times it is called the "hot spot") which mediate the high energy interaction.

At a low multiplicities we deal with the events with the only one Pomeron exchange
 (for simplicity here we will call this object - the Pomeron ). Thus at a 
low multiplicity the BEC have measured the size of one individual Pomeron. Looking at the high multiplicity evens we select the configuration with the many Pomerons. Combinatorically dominates the contribution where the pions are radiated by two different Pomerons. In this case the radius, $r$, measured by BEC characterized the mean separation between the Pomerons. It should be larger than that in a low multiplicity events. Such a growth of $<r>$ with $N_{ch}$ was observed both at a lower (CERN-ISR)~\cite{ua1bec} and at the LHC energies~\cite{CMS2}. It was predicted in~\cite{smrk} that at a larger multiplicities, when the number of the Pomerons become very large and these objects populate all the area where the two incoming proton overlap, there sholud be the "saturation" of the radius measured by BEC, that is $<r(N_{ch})>\rightarrow const$. Such a saturation, indeed, was observed in the ATLAS experiment (see Fig.3 of the recent paper~\cite{atlas}).  Moreover the value of this constant ($\simeq 2.3$ fm) is in a good agreement with that, $const\simeq 2.2$ fm evaluated in ~\cite{smrk} based on the $t$-slope of elastic $pp$-scattering $B_{el}=20$ GeV$^{-2}$.\\

\section{Outlook}
It would be interesting to study the BEC in more details selecting the different sorts of events. In particular, one may compare the radiuses measured in pion-pion and kaon-kaon correlations.

Another possibility is to study the correlations between two pions with relatively large transverse (with respect to the beam direction) momenta $k_T\sim 1\ -\ 2$ GeV. At $k_T=1$ GeV the inclusive cross section $d\sigma/d^3k$ falls down more than two orders of magnitude~\cite{cms-t}. So there is small probability that these two pions with  relatively large $k_T$ flying in the same direction (in order to have a small $Q$ corresponding to the peak in Bose-Einstein correlation) are radiated by two different Pomerons. The dominant contribution comes from the situation where two such pions are produced in the fragmentation of the same (mini)jet.

Recall that from the viewpoint of hadronization the pions originated by the Pomeron are produced in a non-perturbative regime by some 'tube' (or 'colour string') which transfer the antisymmetric colour octet flux from the incoming proton to that in the opposite beam (at least in the case of the LO BFKL Pomeron it is the 
antisymmetric colour octet flow). In the case of a high $E_T$ jets at the LHC they are  mainly the gluon jets which again forms the antisymmetric colour octet tube. That is it would be reasonable to expect the saturation of the value of $<r(k_T)>\to  const$ (now measured as the function of the quark transverse momenta) at more or less the same value as that measured in a low multiplicity (i.e. one Pomeron) events since in both cases the BEC have measured the size of the (almost) the same colour tube. The decrease of the radius with $k_T$ increases from  $k_T\sim 0.2$ GeV to $k_T\sim 1$ GeV was already observed at the LHC (see e,g,~\cite{CMS2,atlas,alicepp}). Now we expect that this decrease should be stopped and in the next interval $k_T=1\ - 2$ GeV the value of $<r>$ should be approximately constant and practically does not depend on the particle density, $N_{ch}$, observe in this particular event.

\section*{Acknowledgements}

The work of MGR  was supported by the RSCF grant 14-22-00281.

\end{document}